# Revisiting Chien-Hrones-Reswick Method for an Analytical Solution

Senol Gulgonul

*Electrical and Electronics Engineering, Ostim Technical University, Ankara Turkey*
*senol.gulgonul@ostimteknik.edu.tr*

**Abstract:** This study presents an analytical method for tuning PI controllers in First-Order with Time Delay (FOTD) systems, leveraging the Lambert W function. The Lambert W function enables exact pole placement, yielding analytical expressions for PI gains. The proposed approach identifies a critical condition that achieves a step response without overshoot with minimum settling time, while also providing explicit tuning rules for systems where controlled overshoot is specified. The method demonstrates strong agreement with established empirical Chien-Hrones-Reswick tuning rules for both non-overshooting and overshooting cases, bridging the gap between theoretical analysis and empirical results.

## 1. Introduction

Industry surveys indicate Proportional-Integral-Derivative (PID) controllers dominate process control applications, with 98% of regulatory loops in refining, chemical, and pulp/paper industries employing this algorithm [1]. Most of these PID controllers are PI and derivative action is not used very often due to accuracy and noise of measurement sensors which requires filtering for a healthy use of derivative [2].

Ziegler and Nichols (1942) established the first PID tuning methodology by experimentally determining that a stable control could be achieved by targeting a 25% amplitude ratio in the closed-loop step response [3]. Ziegler-Nichol's reaction curve defined a first-order with time delay (FOTD) system. The Cohen-Coon method established an analytical foundation for Ziegler-Nichols' empirical tuning rules by employing pole placement techniques while preserving the quarter-amplitude decay stability criterion [4].

Chien, Hrones, and Reswick (CHR) conducted systematic empirical studies on FOTD systems, publishing PID tuning tables for both no-overshoot and 20% overshoot responses. Minimization of settling time without overshoot was an important design criteria of CHR method. [5].

Lambda method, which requires only one tuning parameter ($\lambda$), first suggested by Dahlin (1968), is a pole-placement method designed for FOTD plants characterized by a time constant $T$ and delay $L$ [6].

Increased processing power allowed to use optimization methods for PID tuning. In fact, PID tuning can be defined as a constrained optimization, calculating of PID coefficients for desired performance requirements for the given plant system. Besides fundamental performance requirements like settling time, percent overshoot, gain margin, phase margin, robustness, load disturbance attenuation and measurement noise response have to be considered for optimization [7].

Recent methods based on the Lambert W function have enabled analysis and control of Linear Time Invariant (LTI) Time Delay Systems (TDS) with approaches analogous to Ordinary Differential Equation (ODE) systems [8]. Use of Lambert W function is advantageous compare to first order or Padé approximations to time delay terms.

This study revisits the CHR tuning method by employing the Lambert W function to solve the characteristic equation of a closed-loop system comprising a PI controller and a FOTD plant. This work derives closed-form expressions for PI controller coefficients using the Lambert W function, achieving two key advances: (1) rigorous analytical justification for the empirical tuning rules of the CHR method, and (2) explicit characterization of the trade-off between overshoot and settling time through the dimensionless parameter $\gamma = KK_i eL$.

## 2. Methodology

### 2.1. Analytical Solution for PI Control of FOTD Systems

A first order with time delay (FOTD) plant can be represented as:

$$G(s) = \frac{K}{sT+1} e^{-sL} \tag{1}$$

where K is the process gain, T is the time constant, and L is the time delay.

The PI controller has the form:

$$C(s) = K_p + \frac{K_i}{s} \tag{2}$$

with $K_p$ and $K_i$ as the proportional and integral gains, respectively.

Closed loop system transfer function $G_{cl}(s)$ is derived as:

$$G_{cl}(s) = \frac{C(s)G(s)}{1+C(s)G(s)} = \frac{(K_p + \frac{K_i}{s})\frac{K}{sT+1}e^{-sL}}{1+(K_p + \frac{K_i}{s})\frac{K}{sT+1}e^{-sL}} \tag{3}$$

Simplifying numerator and denominator:

$$G_{cl}(s) = \frac{K(K_p s + K_i)e^{-sL}}{s(1+sT) + K(K_p s + K_i)e^{-sL}} \tag{4}$$

To simplify the analysis, we impose the relation $K_p = TK_i$ which was determined through numerical optimization to minimize settling time for zero overshoot. Substituting $K_p = TK_i$ yields:

$$G_{cl}(s) = \frac{KK_i(1+sT)e^{-sL}}{s(1+sT) + KK_i(1+sT)e^{-sL}} \tag{5}$$

Canceling $(1 + sT)$ from numerator and denominator will further simplify the transfer function:

$$G_{cl}(s) = \frac{KK_i e^{-sL}}{s + KK_i e^{-sL}} \tag{6}$$

The poles of Gcl(s) satisfy the characteristic equation:

$$s + KK_i e^{-sL} = 0 \tag{7}$$

Rearranging terms and multiplying by $Le^{sL}$, we obtain:

$$sLe^{sL} = -KKiL \tag{8}$$

Let $z = sL$ and $\gamma = KK_i eL$ The equation becomes:

$$ze^z = -\gamma/e \tag{9}$$

The solutions are given by the Lambert W function:

$$z = W_k\left(-\frac{\gamma}{e}\right), \quad k \in Z \tag{10}$$

where $W_k$ denotes the *k*-th branch of the Lambert W function. For real-world implementations, only the principal branch (k=0) and first negative branch (k=−1) are typically needed, as higher branches contribute negligibly to dynamics. The poles are then expressed as:

$$s = \frac{W_k\left(-\frac{\gamma}{e}\right)}{L} \tag{11}$$

Three distinct cases emerge based on the value of $\gamma = KK_i eL$:

1. **Critically Damped ($\gamma = 1$)**

   When $\gamma = 1$, the argument of the Lambert W function becomes $-1/e$. Here, the primary and secondary branches of W coincide at $z = -1$, resulting in a double real pole:

   $$s_{12} = \frac{W_0\left(-\frac{1}{e}\right)}{L} = \frac{W_{-1}\left(-\frac{1}{e}\right)}{L} = -\frac{1}{L} \tag{12}$$

   This corresponds to the fastest non-oscillatory response with zero overshoot.

2. **Overdamped ($\gamma < 1$)**

   For $\gamma < 1$, the equation $ze^z = -\gamma/e$ admits two distinct real solutions on the $W_0$ and $W_{-1}$ branches. The poles are:

   $$s_1 = \frac{W_0\left(-\frac{\gamma}{e}\right)}{L}, \quad s_2 = \frac{W_{-1}\left(-\frac{\gamma}{e}\right)}{L} \tag{13}$$

   both of which are negative and real, leading to an overdamped response.

3. **Underdamped ($\gamma > 1$)**

   When $\gamma > 1$, the solutions become complex conjugates:

   $$s_1 = \frac{W_0\left(-\frac{\gamma}{e}\right)}{L}, \qquad s_2 = s_1^* = \frac{W_{-1}\left(-\frac{\gamma}{e}\right)}{L} \tag{14}$$

   with $Re(s) < 0$. The system exhibits oscillatory behavior with an overshoot determined by the value of $\gamma$.

By using $K_p = TK_i$ and $\gamma = KK_i eL$ equations, we have reached an analytical solution for PI controller of FOTD systems:

$$K_p = \frac{\gamma T}{KeL} \tag{15}$$

$$K_i = \frac{\gamma}{KeL} \tag{16}$$

## 3. Simulation Results

No analytical relation was found for peak overshoot and settling time. Having poles and approximating with 2$^{nd}$ order transfer function did not generate close results to exact simulation results. Thus, we simulated for PI controller results for the range $0.1 < \gamma = KK_i eL < 2.0$ covering critically damped, overdamped and underdamped cases.

### 3.1. Settling Time

The minimum settling time with zero overshoot condition occurs at $\gamma = KK_i eL = 1$ as expected. If $\gamma = KK_i eL = 1$ closed loop transfer function becomes only a function of delay $L$:

$$G_{cl}(s) = \frac{(1/eL)e^{-sL}}{s + (1/eL)e^{-sL}} \tag{17}$$

From the simulations we observed that settling time for critically damped case can be formulated as:

$$T_s = 6.56L \tag{18}$$

If a better settling time required for the system, some overshoot has to be allowed in tradeoff. For example %1 allowed results $T_s = 4.42$ settling time as shown in Figure 1.

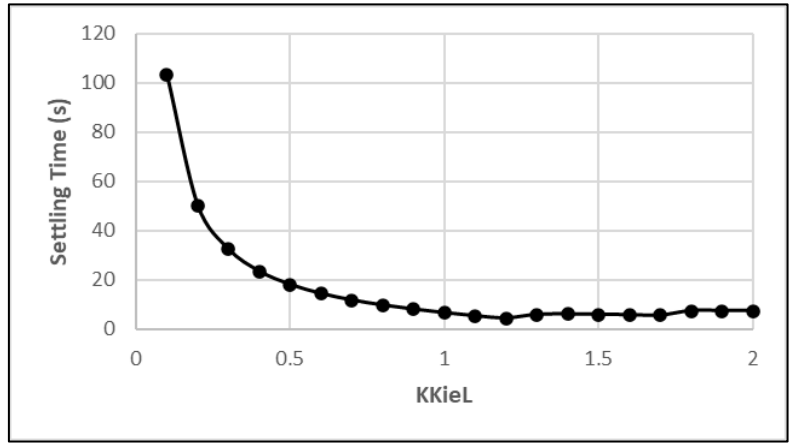

**Figure 1.** Settling time versus $\gamma = KK_i eL$

### 3.2. Peak Overshoot

Peak overshoot is zero for $\gamma = KK_i eL \geq 1$ condition and increases with $KK_i eL$ value as shown in Figure 2.

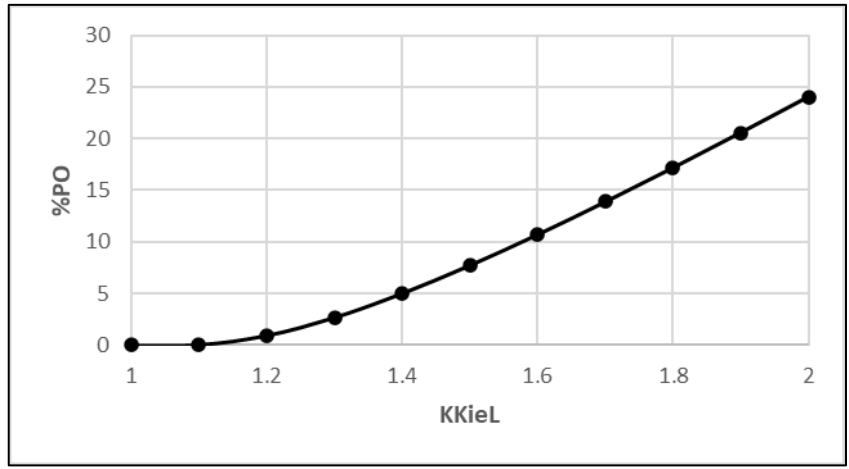

**Figure 2.** Peak overshoot percentage versus $\gamma = KK_i eL$

%20 peak overshoot with $KK_i eL = 1.8837$ value yields PI coefficients very close to CHR method empirical measurements. The step response in Figure 3 demonstrates the 20% overshoot case.

$$K_p = \frac{0.7T}{KL} \tag{19}$$

$$K_i = \frac{0.7}{KL} \tag{20}$$

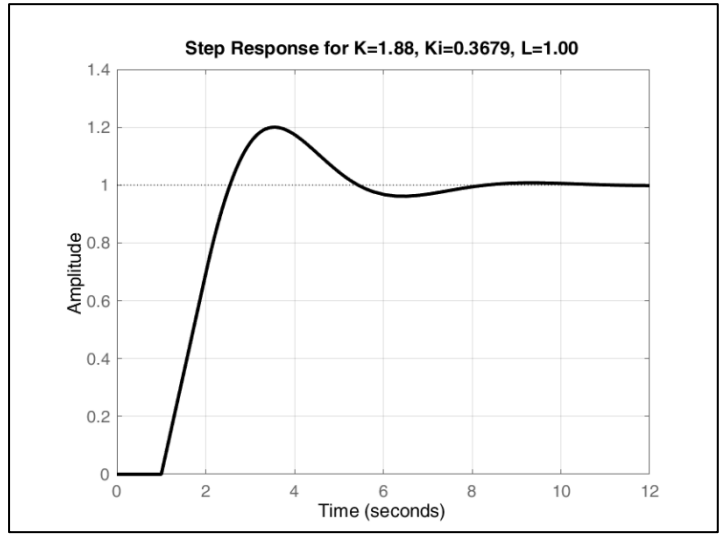

**Figure 3.** Step response for 20% peak overshoot.

## 4. Discussion

PI coefficients for analytical solution and CHR method are summarizes in Table.1. Analytical solution has PI coefficients close to ones empirically measured by CHR method. In analytical solution $\frac{1}{e} = 0.37$ is common multiplier for both $K_p$ and $K_i$ while CHR method has two different multipliers as 0.35 and 0.26 reached empirically.

Using the Lambert W function, we calculated PI coefficients for 20% peak overshoot in order to compare CHR method results. While CHR method has 0.6 empirical multiplier, the analytical solution has multiplier 0.7 to have 20% peak overshoot.

These very close results shows that analytical solution covers empirically measured values in CHR method and provides an additional confidence to the Lambert W solution.

**Table 1.** Comparison of PI coefficients.

|  | No Overshoot | | 20% Overshoot | |
| --- | --- | --- | --- | --- |
|  | $K_p$ | $K_i$ | $K_p$ | $K_i$ |
| **CHR** | $\frac{0.35T}{KL}$ | $\frac{0.29}{KL}$ | $\frac{0.6T}{KL}$ | $\frac{0.6}{KL}$ |
| **Lambert W Solution** | $\frac{0.37T}{KL}$ | $\frac{0.37}{KL}$ | $\frac{0.7T}{KL}$ | $\frac{0.7}{KL}$ |

## 5. Conclusion

The analytical framework developed in this study resolves the PI tuning problem for FOTD systems through Lambert W function. The critical condition $\gamma = KK_i eL = 1$ achieves optimal damping with a settling time proportional to the delay $L$, while $\gamma > 1$ introduces a tunable trade-off between overshoot and speed. Notably, the derived gains for no-overshoot and 20% overshoot ($\gamma = 1.8837$) exhibit remarkable agreement with the empirically validated CHR method, corroborating the analytical model's practical relevance. These results not only generalize classical tuning rules but also provide an analytical approach for designing PI controllers in delay-dominated processes.

**References**

[1] Desborough, L. and Miller, R. (2002) Increasing Customer Value of Industrial Control Performance Monitoring— Honeywell Experience. In: Rawlings, J.B., Ogunnaike, B.A. and Eaton, J.W., Eds., 6th International Conference on Chemical Process Control, AIChE Symp., Series 326, AIChE, New York, 172-192.

[2] Åström, K. J., & Hägglund, T. (2001). The future of PID control. *Control engineering practice*, *9*(11), 1163-1175.

[3] Ziegler, J. G., and Nichols, N. B. (June 1, 1993). "Optimum Settings for Automatic Controllers." ASME. J. Dyn. Sys., Meas., Control. June 1993; 115(2B): 220–222.

[4] Cohen, GHp, and G. A. Coon. "Theoretical consideration of retarded control." Transactions of the American Society of Mechanical Engineers 75.5 (1953): 827-834.

[5] Chien, K. L., Hrones, J., & Reswick, J. B. (1952). On the automatic control of generalized passive systems. *Transactions of the American Society of Mechanical Engineers*, *74*(2), 175-183.

[6] Dahlin E.B., Designing and Tuning Digital Controllers, Instr and Cont Syst, 41 (6), 77, 1968

[7] Åström, K. J., & Hägglund, T. (2006). Advanced PID Control. ISA - The Instrumentation, Systems and Automation Society.

[8] S. Yi, P. W. Nelson, and A.G. Ulsoy, Timedelay Systems:Analysis and Control Using the Lambert W Function, World Scientific, 2010